\documentclass[showpacs,preprintnumbers,superscriptaddress,
amsmath,amssymb]{revtex4}

\usepackage{epsfig}
\usepackage{graphicx}
\usepackage{dcolumn}
\usepackage{bm}
\usepackage{times}

\begin{document}

\title{Modularity-like objective function in annotated networks}

\author{Jia-Rong Xie}
\affiliation{Department of Modern Physics, University of Science and
Technology of China, Hefei 230026, China}
\author{Bing-Hong Wang}\email{bhwang@ustc.edu.cn}
\affiliation{Department of Modern Physics, University of Science and
Technology of China, Hefei 230026, China}

\begin{abstract}
We ascertain the modularity-like objective function whose optimization is equivalent to the maximum likelihood in annotated networks. We demonstrate that the modularity-like objective function is a linear combination of modularity and conditional entropy. In contrast with statistical inference methods, in our method, the influence of the metadata is adjustable; when its influence is strong enough, the metadata can be recovered. Conversely, when it is weak, the detection may correspond to another partition. Between the two, there is a transition. This paper provides a concept for expanding the scope of modularity methods.
\end{abstract}

\maketitle

\section{Introduction}

Community structure, a partition of nodes in which the density of edges within groups is denser than that between groups, is an important large-scale structure in complex networks, and has attracted significant attention in recent years \cite{Fortunato_review,Newman_review,Fortunato_review_16}. Many methods have been proposed for detecting community structure. Here, we focus on two: statistical inference \cite{Decelle_SatInf,Decelle_SatInf2,Karrer_DCSBM} and modularity-based methods \cite{Newman_modularity}. Statistical inference is flexible; it can be used for different purposes, such as detecting generalized communities \cite{Newman_Gen} or estimating group number \cite{Newman_qnum}. Additionally, statistical inference can be used for detecting annotated networks, in which annotations or metadata that describe the attributes of nodes (such as the age, gender, or ethnicity of individuals in a social network) accompany the network structure \cite{Newman_metadata}. Newman-Girvan modularity \cite{Newman_modularity} is the most popular measure of the quality of a partition. Several modifications have been proposed for measuring different unannotated network structures, including weighted \cite{Weighted_mod}, directed \cite{Directed_mod}, bipartite \cite{Bipartite_mod} and multiplex networks \cite{Multiplex_mod}. However, modularity in annotated networks has not been defined. In the paper, we focus on the objective function in these networks and its relation to Newman-Girvan modularity.

The equivalence between modularity optimization and maximum likelihood \cite{Zhang_MBP,Newman_equivalent} may inspire us to our goal. However, this derivation is for unannotated networks. In the statistical inference method, the model of a network with community structure is defined and then fit to observed network data. In most cases, the model parameters are estimated by likelihood maximization; for different considerations or data types, the likelihoods are different. The likelihood in annotated networks differs from (though is similar to) that of unannotated networks. Herein, we ascertain the modularity-like objective function whose optimization is equivalent to the maximum likelihood in annotated networks. We demonstrate that the modularity-like objective function is a linear combination of modularity and conditional entropy. In contrast with the statistical inference method, we set a variable parameter that controls the influence of the metadata. Our results, in both synthetic and real-world networks, demonstrate that if the parameter is strong enough, the metadata can be recovered; however, if it is weak, our method may recover another partition that is more evident, instead of the metadata. Between the two, we find a transition from the more evident partition to the metadata.

\section{Method}

To illuminate our method, we first provide a brief introduction to the likelihood of statistical inference in annotated networks \cite{Newman_metadata}. In this paper, we consider only the case in which the metadata is a classification or a partition of nodes, $\mathbf{x} = \{x_i\}$. In this method, a degree-corrected stochastic block model is defined to a network. The probability, or likelihood, that the model generates a particular network $\mathbf{A}$ and group assignment $\mathbf{s}$ with $q$ groups is

\begin{equation}\label{likelihood}
\begin{split}
P(\mathbf{A},\mathbf{s}|\mathbf{\Theta},\mathbf{\Gamma},\mathbf{x}) = P(\mathbf{A}|\mathbf{\Theta},\mathbf{s}) P(\mathbf{s}|\mathbf{\Gamma},\mathbf{x}) = \prod\limits_{i<j} p_{ij}^{A_{ij}} (1-p_{ij})^{1-A_{ij}} \prod\limits_{i}\gamma_{s_i x_i},
\end{split}
\end{equation}
where $\gamma_{s x}$ is the probability that a node is assigned to group $s$ given its metadata $x$; $\mathbf{\Gamma}$ denotes the matrix of parameters $\gamma_{s x}$; $p_{ij} = k_i k_j \theta_{s_i s_j}$ is the probability of node $i$ connecting to $j$, where $k_i$ ($k_j$) is degree of node $i$ ($j$) and $\theta_{st}$ are parameters indicate the strength of connection between groups; and $\mathbf{\Theta}$ denotes the matrix of parameters $\theta_{st}$.

The likelihood maximization is equivalent to the maximization of the logarithm

\begin{equation}\label{log_likelihood}
\begin{split}
\log P(\mathbf{A},\mathbf{s}|\mathbf{\Theta},\mathbf{\Gamma},\mathbf{x}) &\sim \sum\limits_i \log \gamma_{s_i x_i} + \frac{1}{2} \sum\limits_{ij} A_{ij} \log (k_i k_j \theta_{s_i s_j}) + \frac{1}{2} \sum\limits_{ij} \log(1-k_i k_j \theta_{s_i s_j}) \\
&\sim \sum\limits_x \sum\limits_s N_{sx} \log \frac{N_{sx}}{N_x} + \frac{1}{2} \sum\limits_{ij} A_{ij} \log \theta_{s_i s_j} - \frac{1}{2} \sum\limits_{ij} k_i k_j \theta_{s_i s_j},
\end{split}
\end{equation}
where $N_{sx}$ is the number of nodes assigned to group $s$ with annotation $x$ and $N_x$ is the number of nodes with annotation $x$. The first term is:

\begin{equation}\label{conditional_entropy}
\sum\limits_x \sum\limits_s N_{sx} \log \frac{N_{sx}}{N_x} = N \sum\limits_x \sum\limits_s p(s,x) \log \frac{p(s,x)}{p(x)} = N \sum\limits_x p(x) \left( \sum\limits_s p(s|x) \log p(s|x) \right) = - N H(S|X),
\end{equation}
where $N$ is the number of nodes in the network and $H(S|X)$ is the conditional entropy. The second and third terms induce the modularity \cite{Newman_equivalent}. The planted partition model \cite{Condon_planted} is a special case of the stochastic block model in which the parameters $\theta_{st}$ describing the community structure take only two different values:

\begin{equation}\label{two_values}
\theta_{st} = \left\{
\begin{array}{rcccl}
&\theta_{in}       & & \text{if} &{s = t} \\
&\theta_{out}      & & \text{if} &{s \neq t}
\end{array}
\right. .
\end{equation}
Eq.~(\ref{two_values}) implies that

\begin{equation}
\theta_{st} = (\theta_{in} - \theta_{out}) \delta_{st} + \theta_{out},
\end{equation}

\begin{equation}
\log \theta_{st} = (\log \theta_{in} - \log \theta_{out}) \delta_{st} + \log \theta_{out}.
\end{equation}
Thus, the second and third terms of Eq.~(\ref{log_likelihood}) are \cite{Newman_equivalent}

\begin{equation}
\begin{split}
\frac{1}{2} \sum\limits_{ij} A_{ij} \log \theta_{s_i s_j} - \frac{1}{2} \sum\limits_{ij} k_i k_j \theta_{s_i s_j}
& \sim M\log \frac{\theta_{in}}{\theta_{out}} \frac{1}{2M} \sum\limits_{ij} \left( A_{ij} - \frac{2M (\theta_{in} - \theta_{out})}{(\log \theta_{in} - \log \theta_{out})} \frac{k_i k_j}{2M}\right)\delta_{s_i s_j},
\end{split}
\end{equation}
in which some constants have been dropped. The maximization of Eq.~(\ref{log_likelihood}) is equivalent to the maximization of

\begin{equation}\label{modularity_like}
\frac{1}{2M} \sum\limits_{ij} \left( A_{ij} - \gamma \frac{k_i k_j}{2M}\right)\delta_{s_i s_j} - \alpha H(S|X) = Q(\gamma) - \alpha H,
\end{equation}
where $\gamma = \frac{2M (\theta_{in} - \theta_{out})}{(\log \theta_{in} - \log \theta_{out})}$ and $\alpha = \frac{N}{M(\log \theta_{in} - \log \theta_{out})}$, which can be estimated. In this paper, we set $\gamma = 1$ and treat $\alpha$ as a variable parameter to control the balance between the structure and metadata. High values of $\alpha$ drag the result to the metadata, though the principle for selecting the appropriate value of $\alpha$ is still unknown. We emphasize that our goal is to determine how metadata can be recovered, so the number of groups of detected partitions is equals to that of the metadata in most case. Eq.~(\ref{modularity_like}) is the modularity-like objective function, which is a linear combination of modularity and conditional entropy. We have demonstrated that the optimization of Eq.~(\ref{modularity_like}) is equivalent to the maximum likelihood of Eq.~(\ref{likelihood}). As the modularity-like objective function is known, we use simulated annealing \cite{{Guimera_simuanneal}} for optimization with a fixed $q$.

\section{Results}

\begin{figure}
\scalebox{0.5}[0.5]{\includegraphics{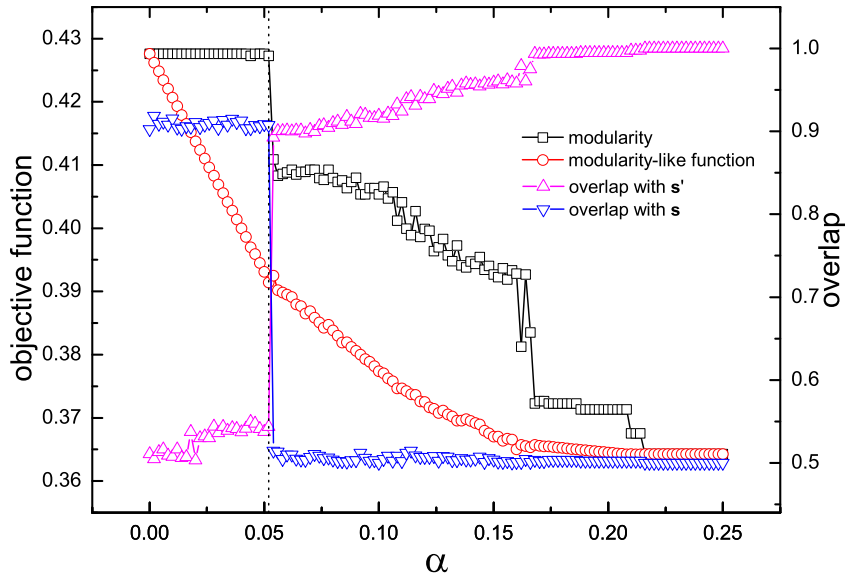}}
\caption{(Color online) The objective functions and overlap of a network generated by the SBM in Eq.~(\ref{q4_SBM}), with $N = 2000$, $c = 3$, $\epsilon_1 = 0.1$ and $\epsilon_2 = 0.15$.}\label{fig:Nested_SBM}
\end{figure}

Our first example is a network generated by a stochastic block model (SBM). In SBM, nodes are randomly assigned to one of $q$ groups and the probability that any pair of nodes connects depends on the node memberships, $p_{ij} = \omega_{s_i s_j}$. In this case, we set $q = 4$ and

\begin{equation}\label{q4_SBM}
\mathbf{\omega} = \frac{4 c} {N (1 + \epsilon_1) (1 + \epsilon_2)}
\left(
\begin{array}{cccc}
 1 & \epsilon_2 & \epsilon_1 & \epsilon_1 \epsilon_2 \\
 \epsilon_2 & 1 & \epsilon_1 \epsilon_2 & \epsilon_1 \\
 \epsilon_1 & \epsilon_1 \epsilon_2 & 1 & \epsilon_2 \\
 \epsilon_1 \epsilon_2 & \epsilon_1 & \epsilon_2 & 1
\end{array}
\right),
\end{equation}
where $c$ is the average degree in the network. It is also a special case of a nested SBM \cite{Peixoto_hierarchical}, in which $L$ (here $L = 2$) community structures are coupled. In the first partition $\mathbf{s}$, original groups 1 and 2 are merged into one group, and the remaining two original groups are merged into another group. In partition $\mathbf{s'}$, original groups 1 and 3 are merged and the left original groups are merged. $\epsilon_1$ and $\epsilon_2$ denote the strength of the two planted structures. In this case, we set $N = 2000$, $c = 3$, $\epsilon_1 = 0.1$, $\epsilon_2 = 0.15$ and metadata $\mathbf{x} = \mathbf{s'}$. $\mathbf{s'}$ is much weaker than $\mathbf{s}$, so that with this metadata, the method in \cite{Newman_metadata} recovers $\mathbf{s}$ rather than $\mathbf{s'}$. However, by adjusting the influence of the metadata with parameter $\alpha$, our method can recover $\mathbf{s}$ in an appropriate range (see Fig.~\ref{fig:Nested_SBM}).

Fig.~\ref{fig:Nested_SBM} shows that the modularity-link function looks like a broken line with three segments. There is a transition at $\alpha_c = 0.052$. Below this transition, $\alpha$ is small enough that structure plays a leading role. Optimization of the objective function finds the partition with the highest modularity. In this case, $Q(\mathbf{s}) > Q(\mathbf{s'})$, so $\mathbf{s}$ is recovered. Above $\alpha_c$, the value of overlaps (i.e., the fraction of nodes correctly detected) with the two structures exchanges. If $\alpha$ is not high, both the structure and metadata play important roles in detection. The metadata provides all information of $\mathbf{s'}$, $H(\mathbf{s'}|\mathbf{x}) = 0$; while it provides no information to $\mathbf{s}$, $H(\mathbf{s}|\mathbf{x})$ is high. Thus, the metadata drags the detection to it. However, the landscape has a smooth valley surrounding $\mathbf{s'}$ \cite{Good_Landscape}. Due to fluctuation, there are some partitions that are correlated with $\mathbf{s'}$ (i.e., the Hamming distance to $\mathbf{s'}$ is low) with higher modularity-like objective functions than those of $\mathbf{s'}$. Optimization methods will recover one of them, so the overlap between the detected partition and metadata is high but not equal to 1. Only when $\alpha$ is high enough, metadata plays crucial role and can be recovered absolutely.

\begin{figure}
\scalebox{0.5}[0.5]{\includegraphics{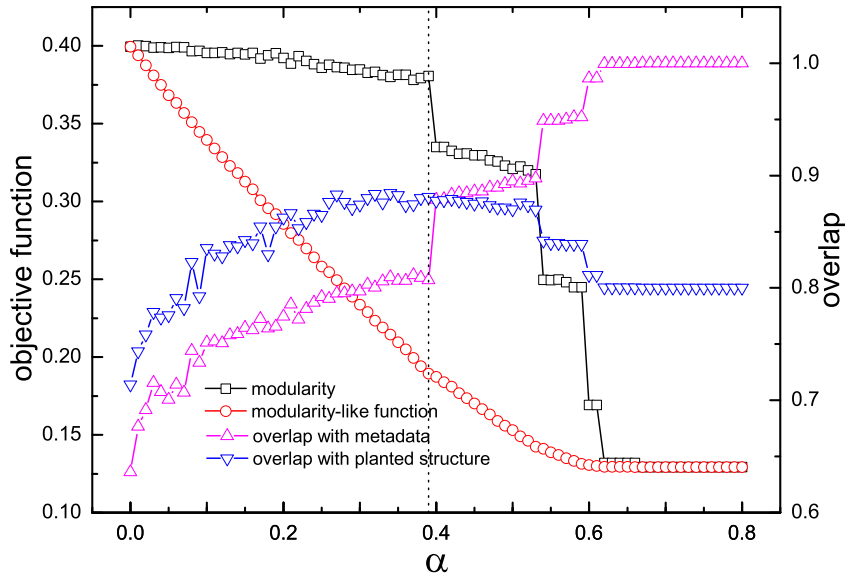}}
\caption{(Color online) The objective functions and overlap of a network generated by a planted partition model with $N = 2000$, $q = 2$, $c = 3$, and $\epsilon = p_{out} / p_{in} = 0.2$.}\label{fig:SBM}
\end{figure}

Our second example is a network generated by a planted partition model, which is a special case of SBM with edge probabilities $p_{in}$ and $p_{out}$ for within-group and between-group edges. We generated node metadata that matched the true planted assignments, but with an error rate of $\rho = 0.2$ to indicate random noise. Without metadata, or if $\alpha = 0$, the approximate planted structure can be recovered. As $\alpha$ increases, detection is gradually dragged to the metadata (see Fig.~\ref{fig:SBM}). If $\alpha$ is high enough, the metadata is recovered absolutely and the overlap with the planted structure was $1 - \rho$. The transition in Fig.~\ref{fig:SBM} is not as strong as that in Fig.~\ref{fig:Nested_SBM}; the overlap with the planted structure in Fig.~\ref{fig:SBM} changes continuously at $\alpha_c$. The planted structure was recovered best at an $\alpha$ value of about $0.34$. In \cite{Newman_metadata}, the strength of the metadata is fixed and may be not the best choice.

\begin{figure}
\scalebox{0.3}[0.3]{\includegraphics{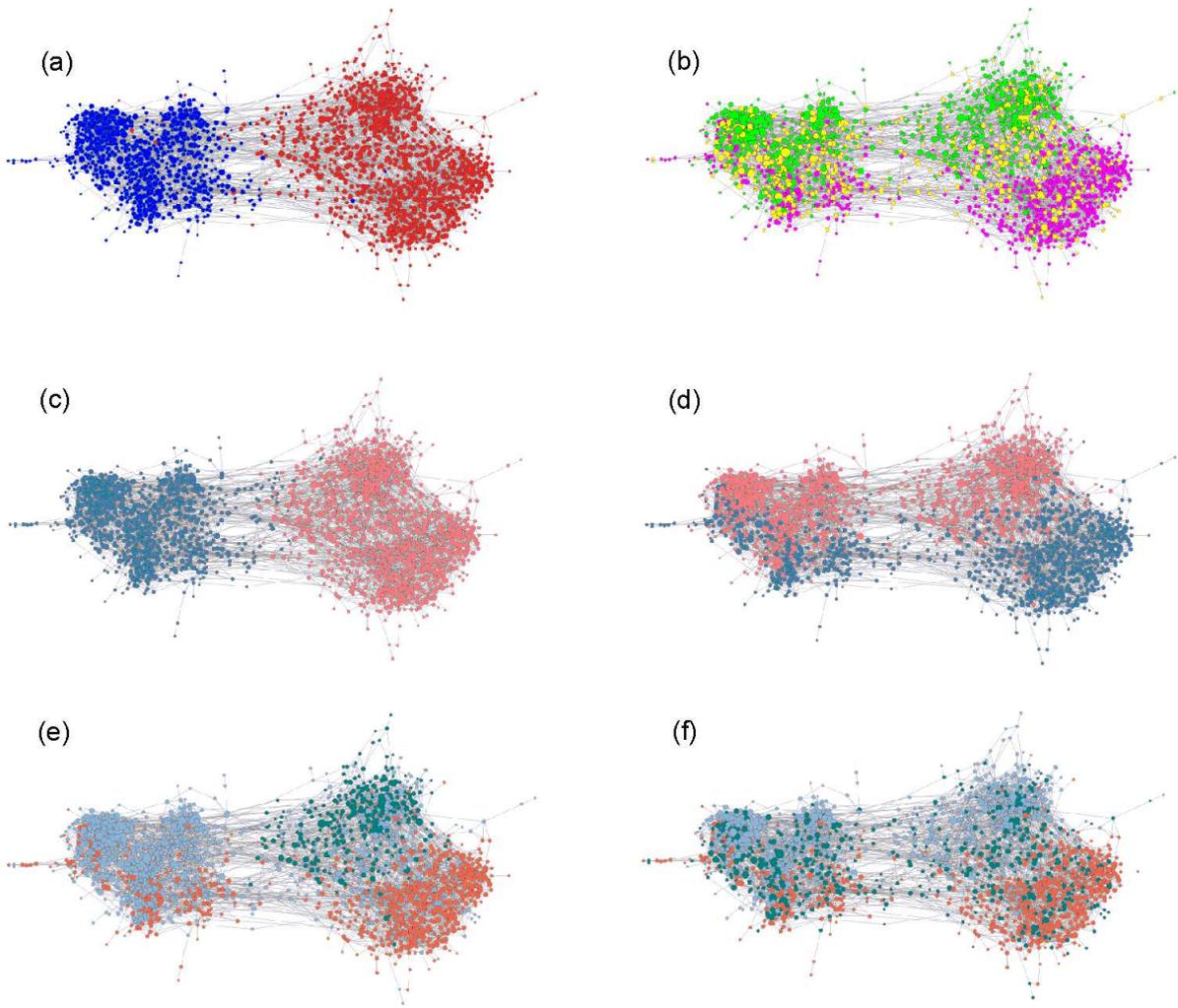}}
\caption{(Color online) The ground-truth and detected partitions in the network of students. (a) The classifications of middle (blue) and high (red) school. (b) Ethnicity metadata: purple for White, green for Black, and yellow for others. (c) The detected partition recovers high/middle school, $q = 2$, $\alpha = 0.1$. (d)-(f) The detected partitions recover ethnicity. (d) $q = 2$, $\alpha = 0.3$, (e) $q = 3$, $\alpha = 0.5$ and (f) $q = 3$, $\alpha = 0.75$. The figures are drawn with the Gephi network visualization software \cite{Bastian_Gephi} and ForceAtlas2 layout algorithm \cite{Jacomy_ForceAtlas2}.}\label{fig:partitions}
\end{figure}

\begin{figure}
\scalebox{0.3}[0.3]{\includegraphics{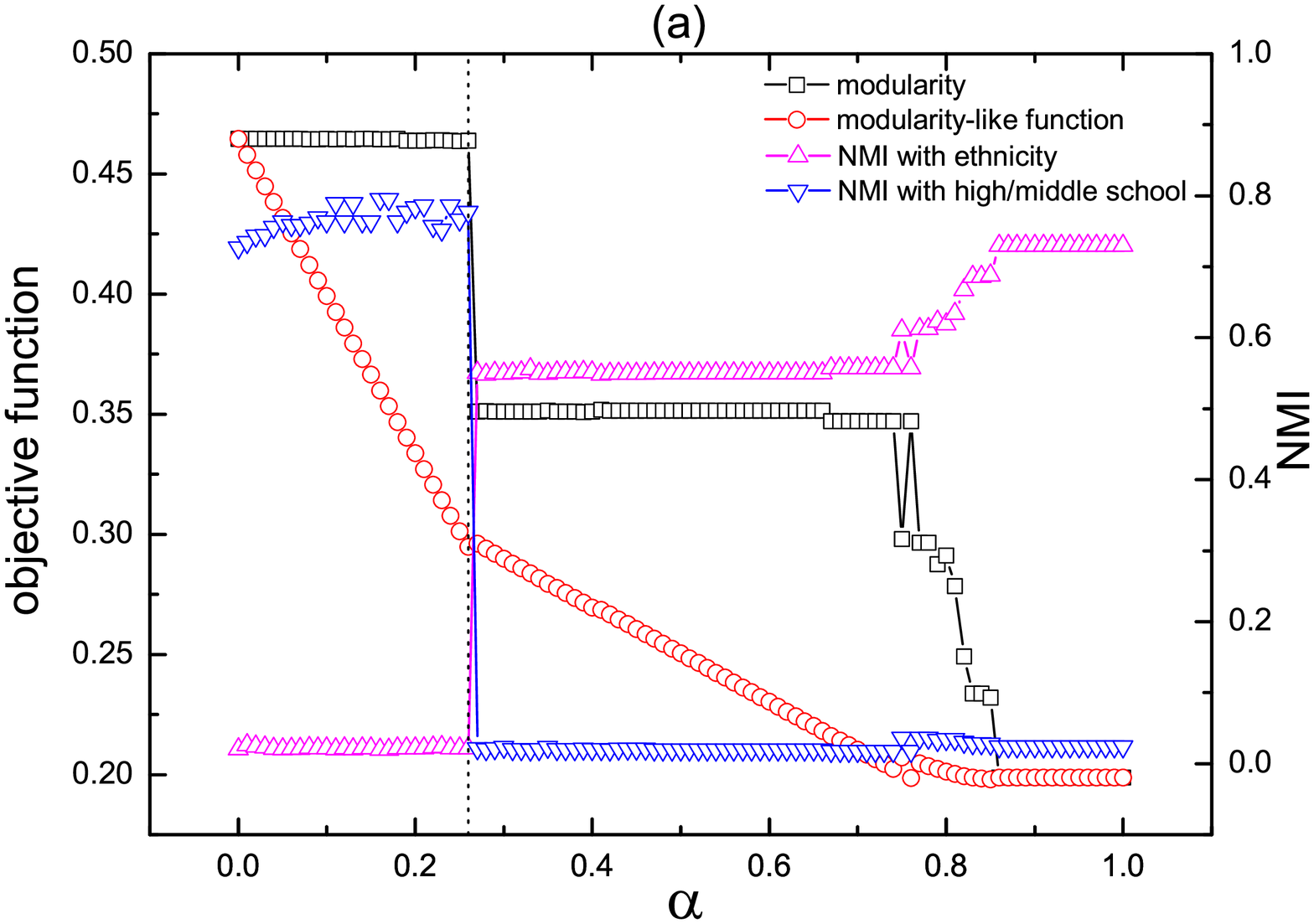}}
\scalebox{0.3}[0.3]{\includegraphics{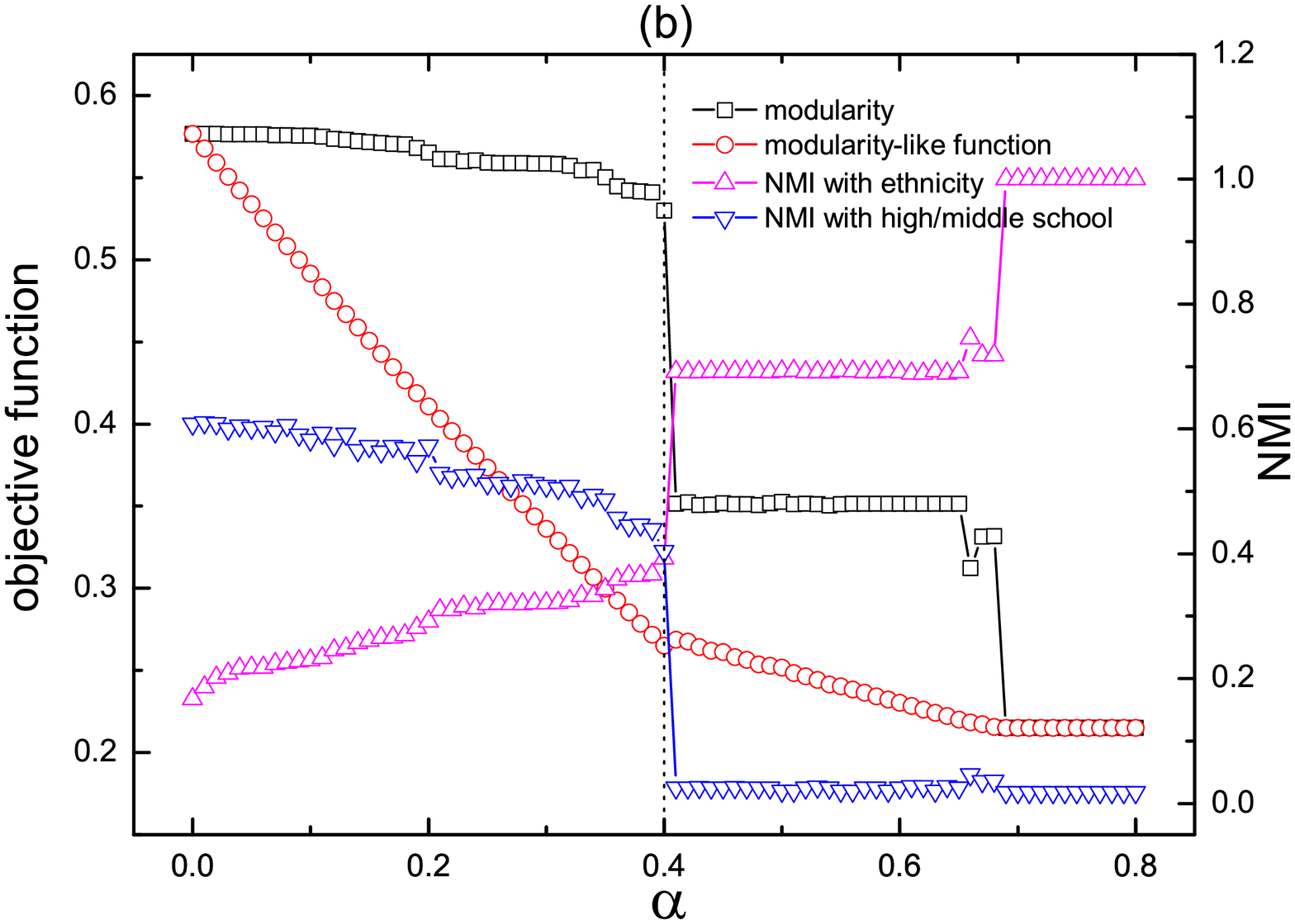}}
\caption{(Color online) The objective functions and NMI of the network of students. (a) With detected group number $q = 2$. (b) $q =3$.}\label{fig:transition}
\end{figure}

Our third example is a network of students drawn from the US National Longitudinal Study of Adolescent to Adult Health \cite{AddHeal}. This network consists of a high school (US grades 9 to 12) and its feeder middle school (grades 7 and 8). The annotations of high/middle school and ethnicity construct two possible partitions (see Fig.~\ref{fig:partitions}(a) and (b)). Between the two, the school is more evident than ethnicity; thus, we treat ethnicity as the metadata. The ethnicity annotation is so weak that with this metadata, the method in \cite{Newman_metadata} recovers the school level rather than ethnicity. However, with $\alpha$, our method can recover ethnicity in an appropriate range (see Fig.~\ref{fig:partitions}(d)-(f) and Fig.~\ref{fig:transition}). Here, we use the normalized mutual information (NMI) \cite{Danon_NMI} rather than overlap to measure how the detected partition matches the annotation, because the detection may have a different group number than the annotations.

\section{Conclusion and discussion}

In this paper, we ascertain the modularity-like objective function whose optimization is equivalent to the maximum likelihood in annotated networks. We demonstrate that the modularity-like objective function is a linear combination of modularity and conditional entropy, with a variable scale $\alpha$ that indicates the influence of the metadata. Unlike in the statistical inference method, our method allows us to adjust the influence of the metadata. Examples in synthetic and real-world networks show that for an appropriate range of $\alpha$ (in which the influence is sufficiently strong), the metadata can be recovered. However, when $\alpha$ is low, another partition may be detected. Between the two values, there is a transition phase.

The statistical inference method is flexible, and it can be used to detect generalized communities \cite{Newman_Gen} and estimate group number \cite{Newman_qnum}. It is therefore interesting to find the corresponding modularity-like objective functions. In this paper, we optimized the modularity-like objective function by simulated annealing. Other optimization algorithms, such as belief propagation \cite{Zhang_MBP}, are left for future work.

\section*{Acknowledgments}

This work is funded by the NSFC (Grant Nos. 11275186, 91024026 and FOM2014OF001).

\end{document}